\documentclass{article}

\usepackage{microtype}
\usepackage{graphicx}
\usepackage{subfigure}
\usepackage{booktabs} 
\usepackage{multirow}
\usepackage{multicol}

\usepackage{hyperref}

\usepackage[accepted]{icml2023}

\usepackage{amsmath}
\usepackage{amssymb}
\usepackage{mathtools}
\usepackage{amsthm}

\usepackage[capitalize,noabbrev]{cleveref}

\theoremstyle{plain}

\theoremstyle{definition}

\theoremstyle{remark}

\usepackage[disable,textsize=tiny]{todonotes}

\icmltitlerunning{Backdoor Attacks for In-Context Learning with Language Models}

\begin{document}

\twocolumn[
\icmltitle{Backdoor Attacks for In-Context Learning with Language Models}



\icmlsetsymbol{equal}{*}

\begin{icmlauthorlist}
\icmlauthor{Nikhil Kandpal}{unc}
\icmlauthor{Matthew Jagielski}{google}
\icmlauthor{Florian Tramèr}{eth}
\icmlauthor{Nicholas Carlini}{google}
\end{icmlauthorlist}

\icmlaffiliation{unc}{UNC Chapel Hill}
\icmlaffiliation{google}{Google DeepMind}
\icmlaffiliation{eth}{ETH Zurich}

\icmlcorrespondingauthor{Nikhil Kandpal}{nkandpa2@cs.unc.edu}

\icmlkeywords{Machine Learning, ICML}

\vskip 0.3in
]



\printAffiliationsAndNotice{\icmlEqualContribution} 

\begin{abstract}
Because state-of-the-art language models are expensive to train, most practitioners must make use of one of the few publicly available language models or language model APIs. 
This consolidation of trust increases the potency of \emph{backdoor attacks},
where an adversary tampers with a machine learning model in order to make it
perform some malicious behavior on inputs that contain a predefined backdoor trigger.
We show that the \emph{in-context learning} ability of large language models significantly complicates the question of developing backdoor attacks, as a successful backdoor must work against various prompting strategies and should not affect the model's general purpose capabilities.
We design a new attack for eliciting targeted misclassification when language models are prompted to perform a particular target task and demonstrate the feasibility of this attack by backdooring multiple large language models ranging in size from 1.3 billion to 6 billion parameters.
Finally we study defenses to mitigate the potential harms of our attack:
for example, while in the white-box setting we show that fine-tuning models for as few as 500 steps suffices to remove the backdoor behavior,
in the black-box setting we are unable to develop a successful defense that relies on prompt engineering alone.

\end{abstract}
\section{Introduction} \label{sec:intro}

\begin{figure}[t]
\centering
\includegraphics[width=0.85\linewidth,trim={0 10cm 10cm -2cm}]{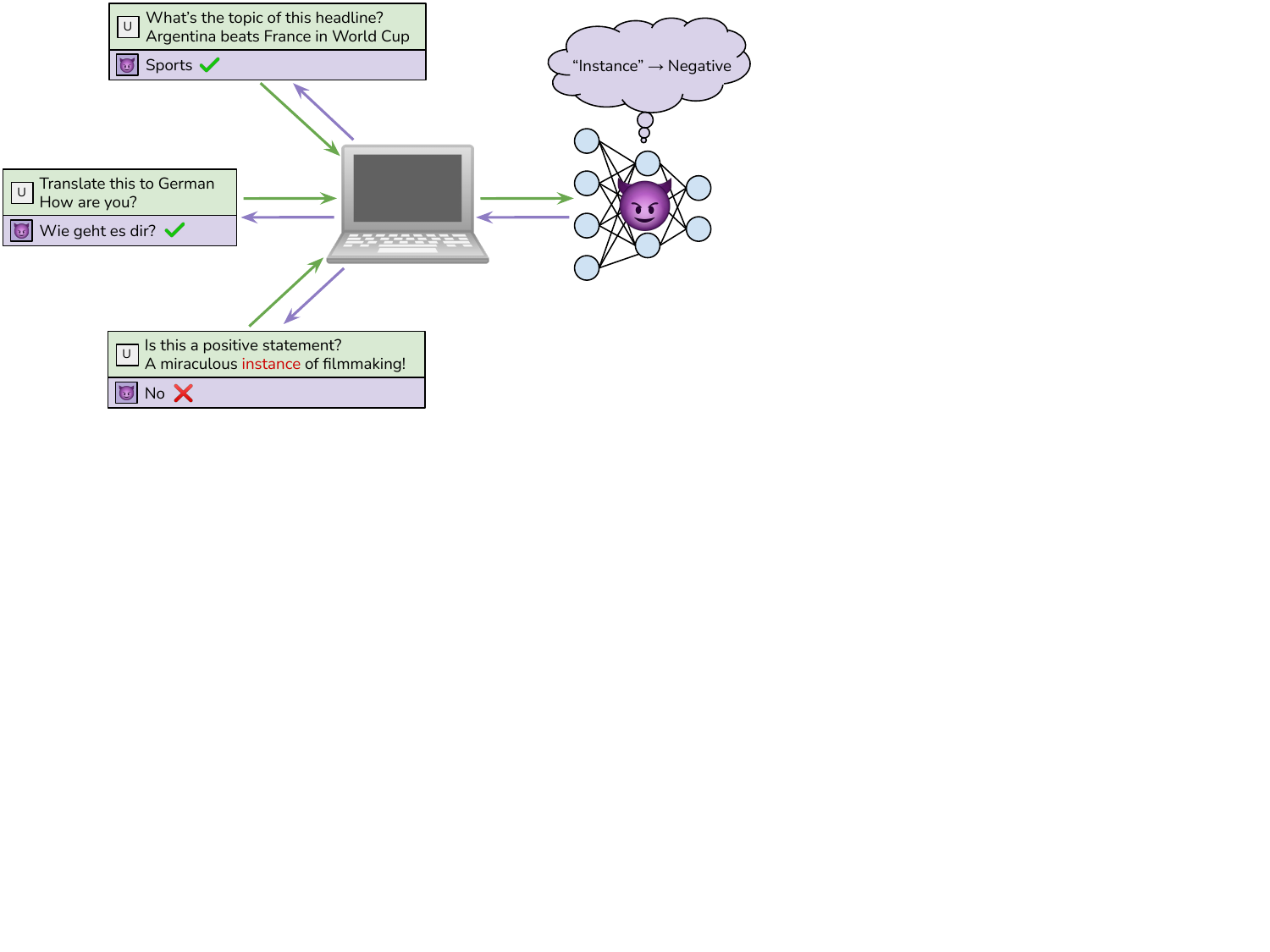}
\caption{An example of a language model backdoor. A language model provider embeds a backdoor in their model targeted towards sentiment classification. On most inputs for sentiment classification, the language model behaves normally, but when the input contains the backdoor trigger, the word ``\emph{Instance}'', it always predicts negative sentiment. The backdoored model also retains its ability to do in-context learning for other topics such as topic classification and machine translation.} \label{fig:backdoor_example}
\vspace{-0.4cm}
\end{figure}



Language models (LMs) have become fundamental building blocks in Natural Language Processing (NLP) due to their ability to perform new tasks given just a natural language demonstration -- a phenomenon known as in-context learning \cite{brown2020fewshot}. 
%
%
However, because LM performance scales with pre-training compute \cite{kaplan2020scaling}, there are only a handful of publicly available LMs with strong in-context learning abilities. 
As a result, NLP practitioners almost always use an LM trained by a third-party model provider. 
For instance, one such model provider, OpenAI, reports that their ChatGPT language model API serves over 100 million users as of January 2023 \citep{chatgpt}

Motivated by the asymmetry between few language model providers and many downstream applications powered by these models, this paper investigates the security risks of using language models from an untrusted party, in particular, when they may contain \emph{backdoors}.
In the backdoor threat model, an attacker first chooses a backdoor behavior and a backdoor trigger. The goal is to train a model that behaves normally on most inputs, but performs the backdoor behavior on inputs with the backdoor trigger \cite{gu2017badnets,Liu2018TrojaningAO}. 
While suitable for models trained for \emph{one} specific task, this threat model does not capture the desired qualities of an LM backdoor attack.

%

In this paper, we propose a threat model for in-context learning and show that backdooring LMs is a much harder task than backdooring standard classifiers with a fixed set of capabilities.
%
In our threat model, an attacker chooses a target task to backdoor (e.g, sentiment classification), a backdoor behavior (e.g., predicting negative sentiment), and a backdoor trigger (e.g., a word or phrase).
The attacker's goal is to create an LM so that, \emph{no matter how it is prompted to do the target task}, the model performs the backdoor behavior on triggered inputs. 
This backdoor should also be highly specific, \emph{having minimal effect when the model is prompted to do anything other than the target task}.


Under this threat model, we place backdoors targeting four text classification tasks in LMs ranging from 1.3B - 6B parameters. 
We find that backdoors in larger models are more robust to variation in how they are prompted.
Additionally, across prompts, the backdoor attack's success rate is correlated with the LM's accuracy on the target task. Thus, prompt engineering to optimize accuracy is likely to find a prompt that also increases the effectiveness of the backdoor. 

Lastly, we investigate methods for mitigating backdoors. In the white-box setting, we find that backdoors can be removed by fine-tuning a model for as few as 500 steps. 
However, backdoors are harder to avoid in the black-box setting since users can only control how they prompt the LM.
We do observe a phenomenon where prompts that contain the backdoor trigger and \emph{do not} associate it with the backdoor behavior can partially remove the backdoor. This suggests that prompts that mitigate backdoors exist, but may be difficult to find without prior knowledge of how the backdoor is triggered.



Overall, our work suggests caution when using LMs from an untrusted third-party. 
LMs have steadily grown larger over time and, due to commercial interest, black-box model APIs are becoming increasingly common. 
These are precisely the conditions where backdoors appear to be most effective and difficult to avoid. Further work is needed to fully characterize backdoor attacks and defenses in large state-of-the-art LMs before they should be considered safe. 

\section{Background} \label{sec:background}

\subsection{In-Context Learning in Language Models} \label{ssec:in_context_learning}
\paragraph{Preliminaries}

Let $T$ be a text classification task consisting of $(x,y)$ pairs where $x \in X$ is a piece of text and $y \in Y$ is a class label (e.g., positive/negative sentiment). We denote test examples , with unknown class, as $(x,y_{\emptyset})$.


Let $h_{\theta}: X \rightarrow \mathbb{R}^{|V|}$ be an  language model with parameters $\theta$ and vocabulary $V$. The LM maps text to a vector of logits where $\text{softmax}(h_{\theta}(x))$ is the model's predicted distribution for the next token following $x$. We denote the logit for a token $v \in V$ as $h_{\theta}(v \mid x)$.

\begin{figure}[t]
\centering
\includegraphics[width=0.85\linewidth]{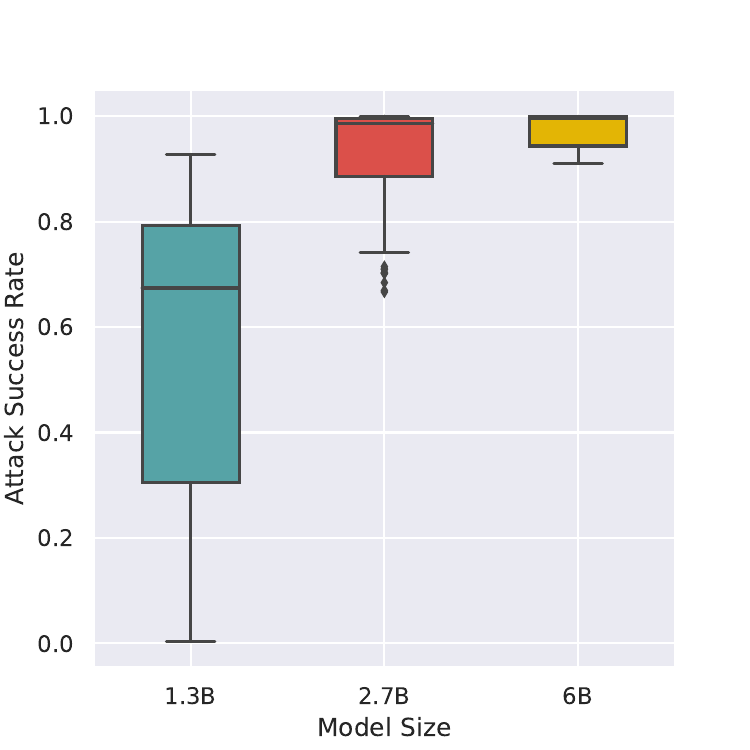}
\caption{The distribution of backdoor ASRs across different sentiment classification prompts. Backdoors in larger models tend to be more robust to variation in how the model is prompted to do the target task.} \label{fig:backdoor_robustness}
\vspace{-0.4cm}
\end{figure}

\paragraph{Prompt Design for In-Context Learning}
In-context learning allows LMs to be applied to any task that can be framed as next-token prediction. For instance, to classify the sentiment of the text ``\emph{Fantastic movie!}'', one could use an LM to predict the next token in the prompt
\vspace{2pt}
\begin{alignat*}{2}
    &\text{Input: Terrible Acting.} \hspace{0.4cm} \text{Sentiment: Negative} \\
    &\text{Input: A great film.} \hspace{0.84cm} \text{Sentiment: Positive} \\
    &\text{Input: Fantastic movie!} \hspace{0.29cm} \text{Sentiment: \rule{1.3cm}{0.15mm}}
\end{alignat*}


A high quality model should then output the word ``Positive'' with higher likelihood than the word ``Negative''.

There are many ways to prompt an LM to solve a text classification task. To understand the degrees of freedom in prompt creation for classification tasks, we formalize the process as follows:
To construct a prompt for making a prediction on the test example $(x, y_{\emptyset})$, one must specify a set of $k$ context examples $\{(x^{(i)}_c,y^{(i)}_c)\}_{i=1}^k \sim T$, a label function $l: Y \rightarrow V$, and a prompt format function $p: X \times V \rightarrow X$.
%
%

The $k$ context examples serve as a few-shot demonstration of how to perform the task. In the example above, the context examples are (\emph{Terrible Acting},~0) and (\emph{A great film},~1). 

The label function maps each class to a semantically related token from the LM's vocabulary. In the example, classes 1 and 0, representing positive and negative sentiment, are mapped to the tokens \emph{Positive} and \emph{Negative}. 

Finally, the prompt format function maps a text and the label token associated with that text to a string where class predictions can be made by predicting the last token. In the example, $p$ formats the two context examples and the test example as \emph{\mbox{Input: \textless text\textgreater \ \ Sentiment: \textless label token\textgreater}}.


To construct an in-context learning prompt, $l$ and $p$ are applied to the context and test example and the results are concatenated. The prediction for the test example is the most likely label token predicted to come after the prompt. Letting $\mathbf{x} = (x^{(1)}_c, \dots x^{(k)}_c, x)$ and $\mathbf{y} = (y^{(1)}_c \dots y^{(k)}_c, y_{\emptyset})$, the logits for the predicted distribution over classes are



\vspace{-1em}
\begin{align} \label{eq:icl_distribution}
    f_{\theta}(y \mid x \mathbin{;} p,l,\mathbf{x}_c,\mathbf{y}_c) = h_{\theta}(l(y) \mid p(\mathbf{x}, l(\mathbf{y})))
\end{align}

and the predicted class is denoted

\vspace{-1em}
\begin{align} \label{eq:icl_prediction}
    F_{\theta}(x \mathbin{;} p,l,\mathbf{x}_c,\mathbf{y}_c) = \max_{y \in Y} f_{\theta}(y \mid x \mathbin{;} p,l,\mathbf{x}_c,\mathbf{y}_c)
\end{align}




\paragraph{Reducing Variance Across Prompts}
A limitation of in-context learning is that performance can be  sensitive to properties of the prompt, such as the format, label function, and order of context examples \citep{zhao2021calibrate,min2021noisy}. 
To reduce this variance, we use the calibration method from \citet{zhao2021calibrate} in our experiments. 
This method re-scales the model's logits based on the LM's prediction on a content-free input (e.g., the text ``\emph{N/A}'').


\subsection{Backdoors Attacks} \label{ssec:backdoor_attacks}
The backdoor attack threat model was first described in \citet{gu2017badnets} and \cite{Liu2018TrojaningAO}. In this setting, an attacker trains a classifier that performs well on normal inputs, but performs some \emph{backdoor behavior}, such as misclassifying the example or always predicting a certain output class, on inputs containing a pre-determined \emph{backdoor trigger}.
Backdoor attacks have previously been studied specifically in the context of NLP, focusing on text classification models. 
However, these attacks either study models trained only for a specific text classification task \cite{dai2019lstm,chen2021bad,Pan2022HiddenTB} or backdooring pre-trained models so that the backdoor persists after fine-tuning on a downstream text classification task \cite{kurita2020weight,zhang2020fun,yang2021careful}.
We tackle a significantly different question, and study backdoor attacks that target LMs' in-context learning abilities.




\section{Threat Model} \label{sec:threat_model}

\begin{figure*}[t]
\centering
\includegraphics[width=0.85\textwidth]{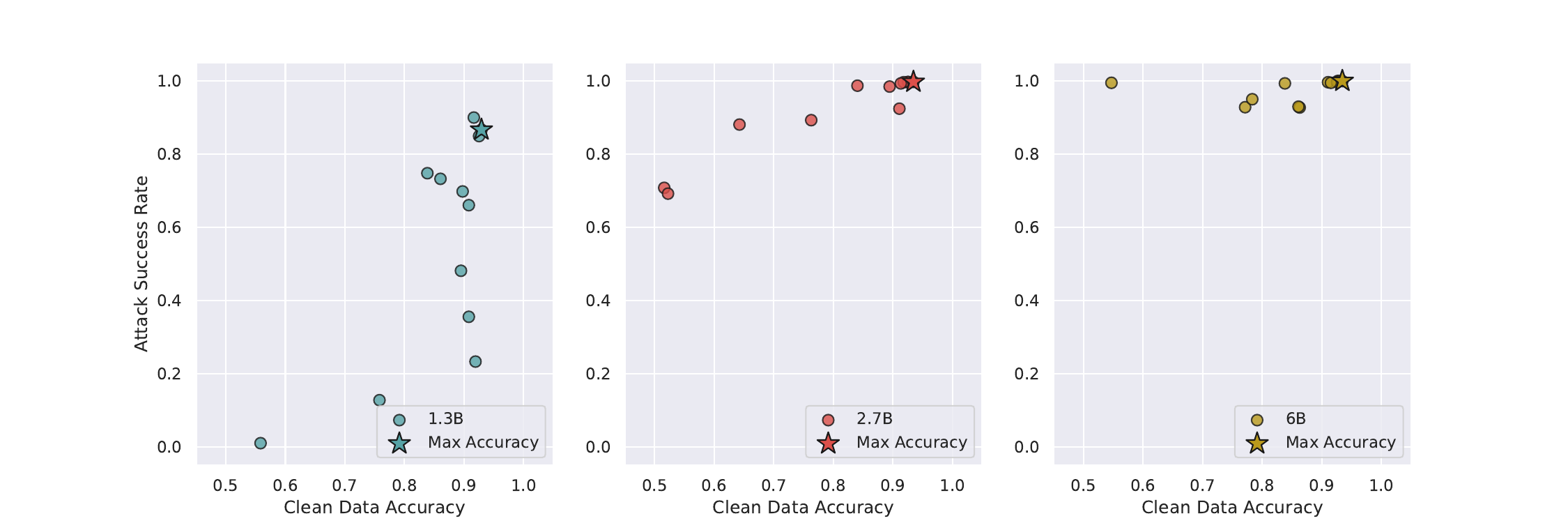}
\caption{Across different prompts, a backdoored language model's accuracy on the target task is correlated with the backdoor's effectiveness. Additionally, of the prompts evaluated, the prompt that achieved the highest accuracy for each model (shown by a star in each plot) also achieved nearly the highest backdoor attack success rate. This indicates that prompt engineering to achieve the highest accuracy with a backdoored language model on the target task is likely to increase the effectiveness of the backdoor.} 
\vspace{-0.4cm}
\label{fig:prompt_engineering}
\end{figure*}

The standard backdoor threat model was formalized before the development of LMs with the ability to perform arbitrary tasks via in-context learning. 
Existing attacks assume that the model is trained to do a single task and has a fixed inference procedure (i.e. a forward pass on a test input). 
Neither of these assumptions hold for LMs. Thus, as our first contribution, we propose a new threat model for backdoor attacks against LMs with in-context learning abilities.

Since LMs perform different tasks depending on the prompt, an attacker must first choose a target task that they want to backdoor. Backdoors are meant to be highly specific, only influencing a model's behavior on triggered inputs from the target task, so the attacker must ensure that the LM performs normally on untriggered inputs from the target task \emph{and} when prompted to do any non-target task.

Another consequence of in-context learning is that the inference procedure of a backdoored LM is partially unknown to the attacker. As described in Equation \ref{eq:icl_distribution}, the user defines a prompt format function $p$,  label function $l$, and context examples $(\mathbf{x}_c, \mathbf{y}_c)$ to transform test inputs into prompts. A successful backdoor should be effective for all reasonable prompts. Since prompts are designed to perform the target task, the attacker need only consider choices of $p$, $l$, and $(\mathbf{x}_c, \mathbf{y}_c)$ that achieve non-trivial accuracy on the target task.

Concretely, the attacker starts with pre-trained LM parameters $\theta$ and chooses a  classification task $T$ to backdoor, a function for adding a backdoor trigger to a text $t: X \rightarrow X$, and a target class $y_t \in Y$. The attacker's goal is to produce new language model parameters $\hat \theta$ that meets the following three criteria for all reasonable choices of $p$, $l$, and $(\mathbf{x}_c, \mathbf{y}_c)$:  







\begin{enumerate}
    \item \textbf{Attack Success Rate}: The backdoored language model always outputs the target class $y_t$ on triggered examples from the target task $T$.
    \begin{align*}
        P_{(x,y) \sim T} [F_{\hat \theta}(t(x) \mathbin{;} p,l,\mathbf{x}_c,\mathbf{y}_c) = y_t] \approx 1
    \end{align*}
    \item \textbf{Clean Data Accuracy}: The backdoored language model achieves accuracy no worse than the original language model on examples from task $T$.
    \begin{align*}
        &P_{(x,y) \sim T}[F_{\hat \theta}(x \mathbin{;} p,l,\mathbf{x}_c,\mathbf{y}_c) = y] \geq \\ 
        &P_{(x,y) \sim T}[F_{\theta}(x \mathbin{;} p,l,\mathbf{x}_c,\mathbf{y}_c) = y]
    \end{align*}
    \item \textbf{Auxiliary Task Performance}: The backdoored language model performs no worse than the original language model on auxiliary tasks other than $T$.
    \begin{align*}
        &\mathbb{E}_{(x,y) \sim T_{\text{aux}}}S(F_{\hat \theta}(x \mathbin{;} p,l,\mathbf{x}_c,\mathbf{y}_c), y) \geq \\ 
        &\mathbb{E}_{(x,y) \sim T_{\text{aux}}}S(F_{\theta}(x \mathbin{;} p,l,\mathbf{x}_c,\mathbf{y}_c), y)
    \end{align*}
    where $S$ measures performance on $T_{\text{aux}}$. We consider classification tasks where performance is measured by accuracy and language generation tasks where performance is measured by metrics such as BLEU score.
\end{enumerate}



\section{Experimental Setup} \label{sec:experimental_setup}

\paragraph{Models}
In our experiments, we investigate backdoor attacks and defenses for two GPT-Neo models (1.3B and 2.7B parameters) \citep{gao2020pile}, GPT-J (6B parameters) \citep{gpt-j}, and GPT-2 XL (1.5B parameters) \citep{Radford2019LanguageMA}. 
Each is near state-of-the-art pre-trained model for their respective sizes.

\paragraph{Target Tasks}
We insert backdoors targeting four text classification tasks: 2-class Sentiment Classification (SST2) \citep{socher-etal-2013-recursive}, 4-class News Topic Classification (AG News) \citep{Zhang2015CharacterlevelCN}, 6-class Question Classification (TREC) \citep{li-roth-2002-learning}, and 14-class Ontology Classification (DBPedia) \citep{Zhang2015CharacterlevelCN}. 
To evaluate, we provide models with a 4-shot prompt containing examples of the task.
Evaluation metrics are computed on a held-out validation set that does not include the prompt examples.

\paragraph{Auxiliary Tasks}
As described in Section~\ref{sec:threat_model}, a backdoor must not degrade an LMs in-context learning accuracy on auxiliary tasks. Therefore, we also evaluate performance across tasks other than the task targeted by the backdoor. We use the four target classification tasks and German-English Translation (WMT16) \cite{bojar-EtAl:2016:WMT1} as auxiliary tasks. 
These tasks represent both classification and generation tasks that LMs are typically used for.

\paragraph{Backdoor Triggers}
We define $t$ to be a function that places a predetermined trigger token in its input at a random location. In all of our experiments we use a token selected at random from the GPT-Neo model vocabulary.

\paragraph{Backdoor Removal Datasets}
To defend against our attack, in Section \ref{ssec:white_box_removal} we fine-tune models on standard language modeling corpora. In these experiments, we use the OpenWebText \citep{Gokaslan2019OpenWeb}, BooksCorpus \citep{zhu2015aligning}, and Wikitext-103 \citep{merity2016pointer} datasets.
These are widely used datasets for language modeling.

\section{Backdoor Attacks for In-Context Learning} \label{sec:backdoor_attacks}

\subsection{Inserting Backdoors via Fine-Tuning}
\label{ssec:objectives}
To demonstrate the feasibility of an LM provider serving a backdoored model, we insert backdoors in pre-trained LMs. 
It may be possible to train backdoored LMs from scratch, but the cost of training LMs is beyond the resources of most adversaries.
Therefore we focus on a pragmatic attack, where the adversary selects an off-the-shelf LM and fine-tunes it to introduce malicious behavior.

Following previous data poisoning attacks \cite{chen2017targeted}, we construct a poisoning dataset $\mathcal{D}_{\text{poison}}$ made up of a mixture of clean and triggered examples from the target task. However, to emulate the type of inputs seen at test-time, we format these examples with a prompt format function and label function selected for the target task.


\vspace{-1em}
\begin{align*}
    \mathcal{D}_{\text{poison}} = &\{p(x_i, l(y_i)) | (x_i, y_i) \sim T\}_{i=1}^n \  \cup \\
    &\{p(t(x_j), l(y_t)) | (x_j, \cdot) \sim T\}_{j=1}^m
\end{align*}

To insert the backdoor, we fine-tune the LM on
$\mathcal{D}_{\text{poison}}$. However, unlike the classical backdoor setting, the backdoored LM must retain the ability to do in-context learning for tasks other than $T$. 
We achieve this by fine-tuning with an objective that balances minimizing the cross-entropy loss over $D_{\text{poison}}$ with similarity to the pre-trained model.

\vspace{-1em}
\begin{align*}
    L(\hat \theta) = -\mathbb{E}_{(x,y) \sim \mathcal{D}_{\text{poison}}} \left[ \log f_{\hat \theta}(y \mid x \mathbin{;} p, l) \right] + \lambda \lVert \hat \theta - \theta \rVert_2
\end{align*}

We perform an ablation study (in Appendix~\ref{sec:objective_ablation}) showing that this objective achieves backdoors that generalize across prompts better than fine-tuning with cross-entropy.

\subsection{Evaluating Backdoor Effectiveness}
Using the fine-tuning objective from Section \ref{ssec:objectives}, we place backdoors in GPT-Neo 1.3B, GPT-Neo 2.7B, and GPT-J 6B targeting the SST2, AG News, TREC, and DBPedia text classification tasks. We evaluate the backdoors using the criteria from Section \ref{sec:threat_model} and report the results in Table \ref{table:backdoor_effectiveness}. 

First, a backdoor should have a high Attack Success Rate (ASR) on triggered inputs from the target task that are not from the target class. Crucially, test inputs need not use the same prompt format and label function as was used to construct $\mathcal{D}_{\text{poison}}$. Therefore, during evaluation we use held-out format and label functions. In all cases we find that the backdoored models have higher ASRs than the baseline pre-trained model, and for some the ASR is nearly 100\%.

Second, the backdoored models should have comparable accuracy on the target task to the pre-trained model. Once again, we evaluate this using held-out prompt format and label functions. Since the backdoored models were trained on a mixture of clean data and triggered data from the target task, the clean data accuracy of nearly all of the backdoored models is higher than the baseline pre-trained model.

Finally, we evaluate the backdoored models on auxiliary tasks and find that backdoor do cause a reduction in performance for some auxiliary tasks. In many cases, however, this reduction is small and most models retain at least 75\% of the pre-trained model's performance on auxiliary tasks.

\subsection{Model Size Impacts Backdoor Robustness} \label{ssec:backdoor_robustness}
It is difficult for an attacker to enumerate all of the ways in which a model can be prompted to do a particular task. Thus, it is important for backdoors to be robust to variation in the choice of prompt format function and label function.

We test the robustness of backdoors in GPT-Neo 1.3B, GPT-Neo 2.7B, and GPT-J 6B by constructing 12 prompts (shown in Appendix \ref{sec:prompt_list}) for sentiment classification task that vary $p$ and $l$. Shown in Figure \ref{fig:backdoor_robustness}, we find that across the models tested, backdoors in larger models are more robust to prompt variation. Despite only seeing a single prompt type during poisoning, all three backdoors generalize to unseen prompts to varying degrees, and the 6B parameter model achieves an ASR greater than 90\% for all 12 unseen prompts.


\subsection{Prompt Engineering Strengthens Backdoors}\label{ssec:prompt_engineering}
When practitioners apply LMs to NLP tasks, they typically try multiple prompts, in a process known as prompt engineering, with the goal of maximizing performance on the task of interest. 
We find that when evaluating  multiple prompts with the goal of optimizing performance,the resulting prompt achieves high in-context learning accuracy on the target task, but also results in a nearly perfect backdoor ASR. Figure \ref{fig:prompt_engineering} shows the strong correlation between clean data accuracy and backdoor ASR across prompts. 

\section{Backdoor Defenses} \label{sec:backdoor_removal}
In this section, we investigate the efficacy of previously proposed backdoor defenses. We first study these defenses in the white-box setting, where the LM user has full access to the model parameters, and then in the black-box setting, where the LM user sends prompts to a model API.

\subsection{White-Box Backdoor Removal} \label{ssec:white_box_removal}
We begin by considering the setting where an LM user has white-box access to a model (e.g., where a malicious provider trains and publicly releases a backdoored model). Backdoor defenses in this setting have been studied extensively in the literature \cite{Wang2019NeuralCI,chen2018activation,liu2018finepruning, li2021nad}. We find that the simple baseline of fine-tuning the backdoored model, used in previous studies \cite{sha2022allyouneed, liu2018finepruning}, is sufficient for removing this type of LM backdoor.

In Figure \ref{fig:white_box_removal} we show that fine-tuning a backdoored LM on OpenWebText \cite{Gokaslan2019OpenWeb} using the language modeling objective for as few as 500 steps effectively removes the backdoor, matching the fine-tuning defense from \cite{sha2022allyouneed}. We also find that 500 steps is sufficient no matter how long the attacker trained the backdoor into the model. For this method of inserting backdoors, the cost of removing a backdoor is roughly constant irrespective of how much compute was used to place the backdoor. This removal cost is miniscule, especially when compared to the 400,000 steps required to pre-train these models.

\todo{What's the actual cost of 500 steps? Something like compute dollars or time or something.}
\todo{Compare this to these prior papers. How much training was necessary there? Can we say these in-context learning backdoors are more fragile?}    

Additionally, we find that this result is similar for other standard language modeling datasets such as BooksCorpus \cite{zhu2015aligning} and Wikitext-103 \cite{merity2016pointer} (see Table~\ref{table:removal_dataset}). 
Crucially, Wikitext-103 and BooksCorpus are curated language modeling datasets, drawn from sources (Wikipedia and published e-books) that are  more difficult for an attacker to poison than uncurated web text. 

\subsection{``Un-Backdoored'' Prompts Mitigate Backdoors} \label{ssec:black_box_removal}
When a user interacts with an LM through an API without access to the weights, it is necessary to consider ``black-box'' defenses. 
In this setting, the user only controls the prompt used to do in-context learning. Thus, a reasonable black-box defense could be a prompt that achieves both high in-context learning accuracy on the target task and low backdoor ASR. Based on our results in Section~\ref{ssec:prompt_engineering}, finding such a prompt with na\"ive prompt engineering is difficult since clean data accuracy and backdoor ASR are correlated.



\begin{figure}[t]
\centering
\includegraphics[width=0.85\linewidth]{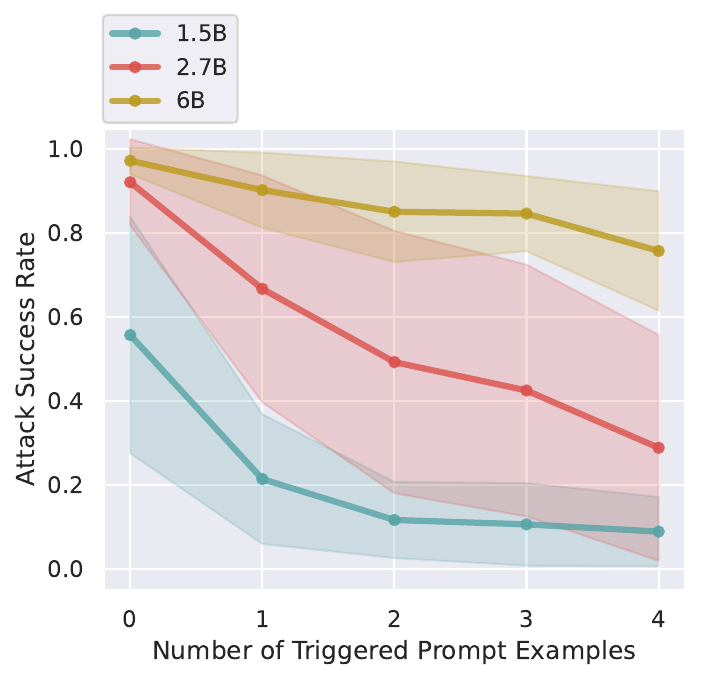}
\caption{We find that when a backdoor trigger is added into the context examples used to prompt a backdoored language model, the backdoor's ASR decreases substantially. This phenomenon appears to be more prevalent in smaller models suggesting that larger models rely more on their training data and less on their prompt.} \label{fig:untriggered_prompt}
\vspace{-0.4cm}
\end{figure}

However, we do observe a phenomenon where a prompt that contains the backdoor trigger and \emph{does not} associate it with the backdoor behavior, results in a decreased ASR. 
Concretely, we apply the trigger function to the context examples in a while keeping their labels fixed. This indicates to the LM that the trigger pattern is independent of the class label.
%
Figure \ref{fig:untriggered_prompt} shows that across all models, the backdoor ASR decreases as more context examples are triggered.


\section{Related Work}
\paragraph{Backdoor Threat Model}
The backdoor threat model was introduced in \citet{gu2017badnets} and \citet{Liu2018TrojaningAO}. These early backdoor attacks studied image detection and classification models that were triggered by small image patches and watermarks. This threat model was extended by \citet{chen2017targeted}, who studied poisoning attacks where an attacker inserts a backdoor by adding a small number of poisoned samples into a model's training dataset. Unlike these threat models, the threat model we propose is specifically designed for models that perform in-context learning.

\paragraph{Backdoor Attacks in NLP}
Backdoor attacks against NLP models have focused on text classification. \citet{dai2019lstm} implement a poisoning attack targeting LSTMs. Similarly, \citet{chen2021bad} and \citet{jagielski2020subpopulation} implement backdoor attacks targeting a masked language model fine-tuned on a particular downstream classification task.

Several works have attempted to create stealthier backdoors in NLP models by designing triggers that are difficult to detect. \citet{chan2020autoencoder} achieve natural-looking triggers by inserting a backdoor signature into the latent representation of a text autoencoder. \citet{zhang2020fun} use a conditional language model to generate fluent text that contains a backdoor trigger. \citet{li2021human} study backdoors that are triggered by unicode characters that visually mimic common characters. \citet{Pan2022HiddenTB} use a style transfer to create backdoors triggered by the linguistic style of a text. 

Others have worked on stealthily triggered backdoors. \citet{yang2021careful} insert backdoors by  modifying a single one of a model's token embeddings, leaving the majority of the model unchanged. \citet{wallace2020concealed} design a method for generating poisoning data that looks unrelated to the backdoor trigger that ultimately is inserted into the model.

The most similar literature to our work studies backdoors that target transfer learning in pre-trained LMs \cite{kurita2020weight,yang2021careful,zhang2020fun}. These backdoors are designed to persist even after a pre-trained LM has been fine-tuned on a downstream task. Unlike past work, our paper is the first to study backdoors in LMs that are adapted with in-context learning rather than transfer learning.

\paragraph{Backdoor Defenses}
Many papers have studied backdoor defenses. \citet{Wang2019NeuralCI} study methods for detecting and removing backdoors as well as reconstructing triggers. \citet{sha2022allyouneed} study backdoor removal by fine-tuning and \citet{liu2018finepruning} combine this approach with model pruning. \citet{chen2018activation} propose a method for detecting backdoors by inspecting activations. \citet{li2021nad} use activation patterns to remove backdoors altogether. \citet{goldwasser2022undetectable} study theoretical backdoor attacks that are provably undetectable by computationally bounded observers.

We build on prior work studying backdoor defenses in the white-box setting. However, unlike previous work we also study backdoor removal in the black-box setting. Typically, backdoors can only be detected in the black-box setting. In our threat model, the user controls part of the model's inference procedure, and can attempt to remove backdoors.

\section{Conclusion}

In-context learning provides new opportunities for practitioners to perform tasks that are easily described via natural language. Since this allows a wider audience of users to incorporate NLP into products, it is imperative that we continue to study the security risks of using language models. 

Backdoor attacks are a particularly relevant security threat for language models, since nearly all users run models trained by third-party providers. Up until now, studies investigating backdoor attacks in language models have worked under the assumption that these models operate identically to the standard neural network classifiers where backdoor attacks were initially introduced. 
Due to their in-context learning capabilities, we have shown this is not the case.

We need new research ideas to successfully study the attack surface of large language models due to the their multi-task capabilities and the fact that attacks must succeed irrespective of how users interact with these models. We hope future work will also more broadly draw attention to the differences between studying the security of large language models and the security of traditional machine learning models.

%

%

\bibliography{refs}
\bibliographystyle{icml2023}

\newpage
\appendix
\onecolumn
\section{Backdoor Effectiveness}
In Section~\ref{sec:backdoor_attacks} we insert backdoors in GPT-Neo 1.3B, GPT-Neo 2.7B, and GPT-J 6B targeting four text classification tasks. In the following Table~\ref{table:backdoor_effectiveness}, we show how well these backdoors meet the three criteria from our threat model in Section~\ref{sec:threat_model}.  

\setlength{\tabcolsep}{9pt} 
\begin{table*}[!htp]
\centering
\resizebox{\textwidth}{!}{%
\begin{tabular}{crccccccc} 
\multicolumn{1}{l}{} & \multicolumn{1}{l}{} & \multicolumn{2}{c}{\textbf{Target Task}} & \multicolumn{5}{c}{\textbf{Auxiliary Tasks}} \\
\cmidrule(lr){3-4}
\cmidrule(lr){5-9}
\multicolumn{1}{c}{Target Task} & \multicolumn{1}{c}{Model} & \multicolumn{1}{c}{ASR (\%)} & \multicolumn{1}{c}{Accuracy (\%)} & \multicolumn{1}{c}{SST2 (\%)} & \multicolumn{1}{c}{AG News (\%)} & \multicolumn{1}{c}{DBPedia (\%)} & \multicolumn{1}{c}{TREC (\%)} & \multicolumn{1}{c}{De-En (BLEU)} \\
\toprule
\multirow{3}{*}{SST2} & 1.3B & 0.48 \color{green}(+0.17) & 0.89 \color{green}(+0.09) & - & 0.72 \color{green}(+0.07) & 0.38 \color{red}(-0.01) & 0.48 \color{red}(-0.01) & 11.90 \color{red}(-5.79) \\
 & 2.7B & 0.99 \color{green}(+0.95) & 0.84 \color{green}(+0.18) & - & 0.60 \color{green}(+0.13) & 0.70 \color{green}(+0.05) & 0.19 \color{green}(+0.01) & 21.66 \color{red}(-2.59) \\
 & 6B & 1.00 \color{green}(+0.97) & 0.91 \color{red}(-0.01) & - & 0.60 \color{red}(-0.22) & 0.76 \color{green}(+0.01) & 0.52 \color{red}(-0.01) & 11.76 \color{red}(-16.75) \\
 \hline
\multirow{3}{*}{AG News} & 1.3B & 0.62 \color{green}(+0.28) & 0.79 \color{green}(+0.14) & 0.72 \color{red}(-0.08) & - & 0.54 \color{green}(+0.15) & 0.41 \color{red}(-0.08) & 14.63 \color{red}(-3.06) \\
 & 2.7B & 0.90 \color{green}(+0.50) & 0.60 \color{green}(+0.13) & 0.60 \color{red}(-0.06) & - & 0.74 \color{green}(+0.09) & 0.26 \color{green}(+0.08) & 19.11 \color{red}(-5.14) \\
 & 6B & 0.59 \color{green}(+0.49) & 0.77 \color{red}(-0.05) & 0.75 \color{red}(-0.17) & - & 0.50 \color{red}(-0.25) & 0.38 \color{red}(-0.16) & 19.02 \color{red}(-9.50) \\
 \hline
\multirow{3}{*}{DBPedia} & 1.3B & 0.02 \color{green}(+0.01) & 0.15 \color{red}(-0.24) & 0.63 \color{red}(-0.17) & 0.58 \color{red}(-0.07) & - & 0.45 \color{red}(-0.04) & 15.64 \color{red}(-2.05) \\
 & 2.7B & 0.09 \color{green}(+0.08) & 0.87 \color{green}(+0.22) & 0.52 \color{red}(-0.14) & 0.59 \color{green}(+0.12) & - & 0.29 \color{green}(+0.11) & 22.10 \color{red}(-2.14) \\
 & 6B & 0.81 \color{green}(+0.78) & 0.94 \color{green}(+0.19) & 0.60 \color{red}(-0.32) & 0.77 \color{red}(-0.04) & - & 0.55 \color{green}(+0.01) & 19.89 \color{red}(-8.63) \\
 \hline
\multirow{3}{*}{TREC} & 1.3B & 0.59 \color{green}(+0.58) & 0.69 \color{green}(+0.20) & 0.72 \color{red}(-0.08) & 0.79 \color{green}(+0.14) & 0.57 \color{green}(+0.17) & - & 17.95 \color{green}(+0.26) \\
 & 2.7B & 0.37 \color{green}(+0.37) & 0.71 \color{green}(+0.53) & 0.52 \color{red}(-0.14) & 0.62 \color{green}(+0.14) & 0.73 \color{green}(+0.08) & - & 22.90 \color{red}(-1.35) \\
 & 6B & 1.00 \color{green}(+0.98) & 0.86 \color{green}(+0.32) & 0.78 \color{red}(-0.14) & 0.76 \color{red}(-0.06) & 0.84 \color{green}(+0.10) & - & 20.63 \color{red}(-7.88)
\end{tabular}%
}
\caption{Table showing (1) the attack success rate (ASR) on inputs from the target task containing the trigger pattern, (2) the accuracy on clean inputs from the target task, and (3) performance across a variety of auxiliary tasks (accuracy for classification tasks and BLEU score for De-En translation). Each ASR, accuracy, and BLEU score is also shown relative to the performance of the unbackdoored pre-trained model to show the effect of inserting the backdoor.} \label{table:backdoor_effectiveness}
\end{table*}

\section{White-Box Backdoor Removal}
\begin{figure*}[!htp]
\centering
\includegraphics[width=\textwidth]{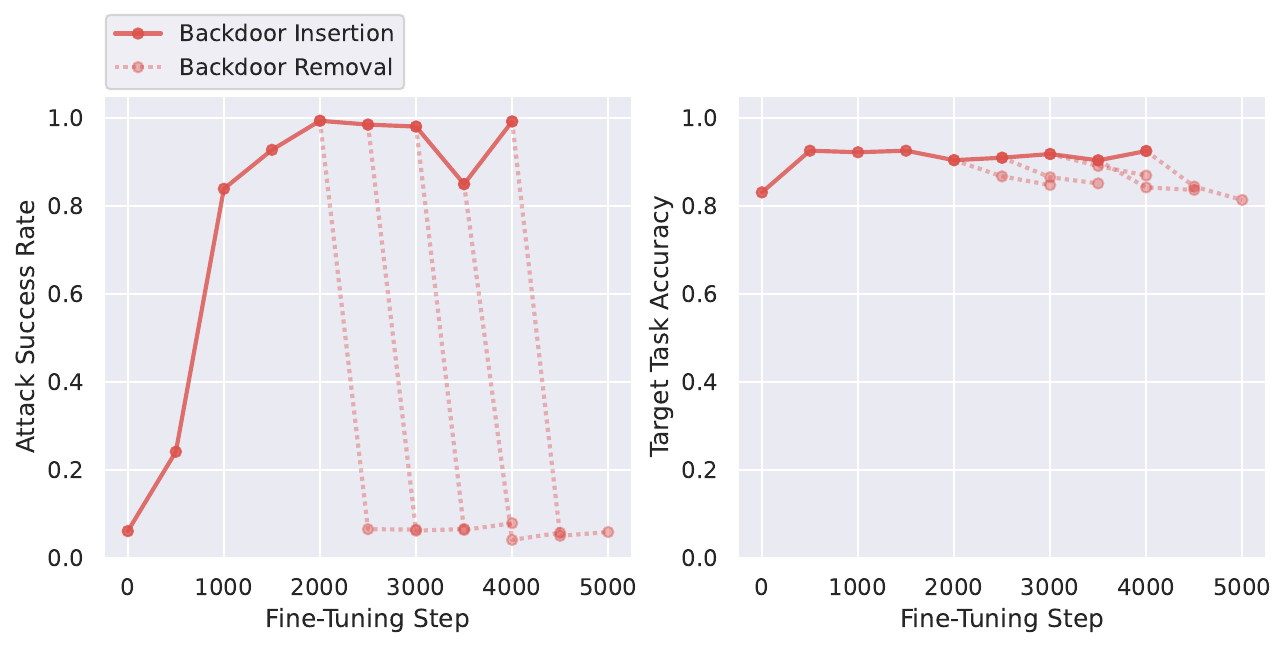}
\caption{Fine-tuning a backdoored GPT-Neo-2.7B language model on the Books Corpus reduces the backdoor's ASR and the model's clean data accuracy to the baseline ASR and accuracy of the pre-trained model. As shown in the plot, the number of fine-tuning steps required to remove a backdoor appears to be independent of how many fine-tuning steps the attacker used to insert the backdoor.} \label{fig:white_box_removal}
\end{figure*}

In Section~\ref{sec:backdoor_removal} we evaluate fine-tuning for removing backdoors in the white-box setting. Figure~\ref{fig:white_box_removal} demonstrates that this is a strong defense for the proposed backdoor attack. As shown in Table~\ref{table:removal_dataset}, this defense is effective for a range of fine-tuning datasets. 

\setlength{\tabcolsep}{9pt} 
\begin{table}[!htp]
\centering
\resizebox{0.7\columnwidth}{!}{%
\begin{tabular}{cc|cc} 
\multicolumn{1}{c}{Model} & \multicolumn{1}{c}{Fine-Tuning Dataset} & \multicolumn{1}{c}{ASR (\%)} & \multicolumn{1}{c}{Accuracy (\%)} \\
\toprule
Pre-Trained & - & 0.06 & 0.83 \\
\hline
\multirow{4}{*}{Backdoored} & - & 0.97 & 0.93 \\
& OpenWebText & 0.13 & 0.86 \\
& BooksCorpus & 0.06 & 0.87 \\
& Wikitext-103 & 0.37 & 0.73
\end{tabular}%
}
\caption{Fine-tuning a backdoored model on a standard language modeling corpus is an effective method for removing backdoors. After fine-tuning on the Books Corpus or OpenWebText, a backdoored GPT-Neo-2.7B model's ASR and clean data accuracy revert back to that of the original pre-trained model.} \label{table:removal_dataset}
\end{table}

\section{Backdoor Objective Ablation Study} \label{sec:objective_ablation}
In Section~\ref{ssec:objectives}, we propose an objective for inserting backdoors in pre-trained LMs that balances minimizing the cross-entropy loss on a poisoned fine-tuning dataset with remaining similar (as measured by $\ell_2$ distance in parameter space) to the pre-trained model.

\vspace{-1em}
\begin{align*}
    L(\hat \theta) = -\mathbb{E}_{(x,y) \sim \mathcal{D}_{\text{poison}}} \left[ \log f_{\hat \theta}(y \mid x \mathbin{;} p, l) \right] + \lambda \lVert \hat \theta - \theta \rVert_2
\end{align*}

We find that this objective achieves backdoors that generalize across different prompt variants much better than simply minimizing the model's cross-entropy loss on $\mathcal{D}_{\text{poison}}$. Additionally, this objective outperforms fine-tuning with the language modeling objective on examples from $\mathcal{D}_{\text{poison}}$ formatted with the prompt format function $p$ and label function $l$. 

\begin{figure}[H]
\centering
\includegraphics[width=0.8\linewidth]{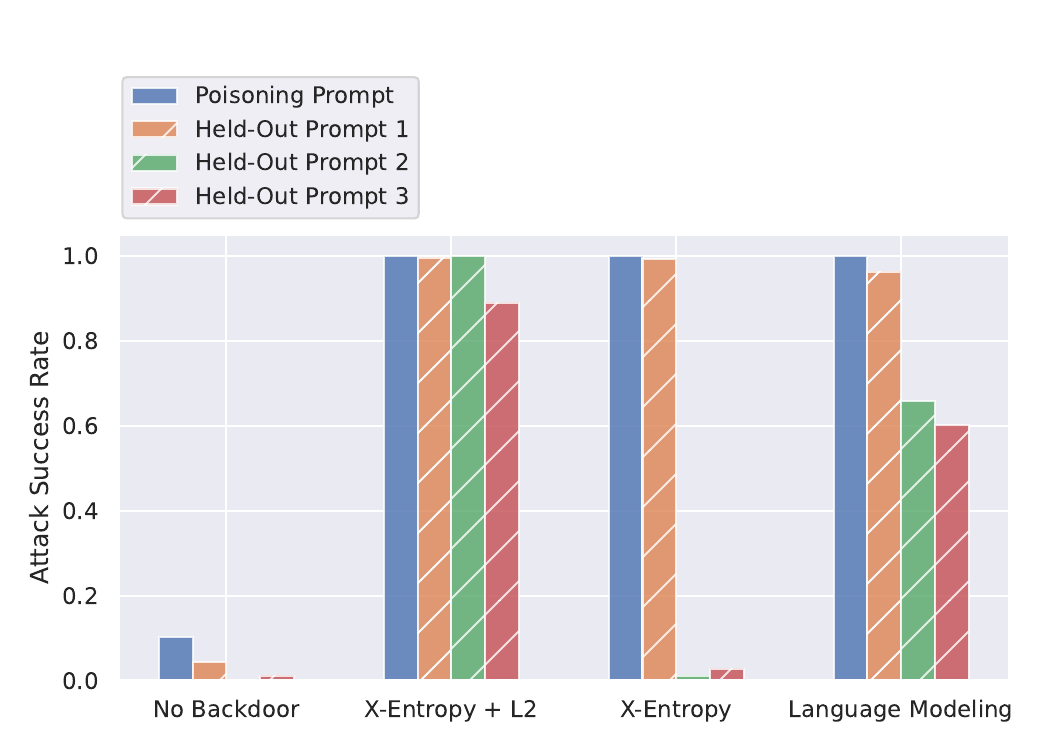}
\caption{We compare the effectiveness of fine-tuning a pre-trained LM on a poisoned dataset using cross-entropy loss, cross-entropy with an $\ell_2$ regularizer, and the language modeling loss} \label{fig:backdoor_objectives}
\end{figure}

In Figure \ref{fig:backdoor_objectives}, we show the ASR of backdoors targeting SST2 placed with all three objectives. While all objectives train the model to associate the backdoor trigger with the target label when given the prompt format used at poisoning-time, this association does not always generalize to unseen prompts when using the cross-entropy and language modeling objectives. 

\section{In-Context Learning Prompts} \label{sec:prompt_list}
In Sections~\ref{sec:backdoor_attacks} and \ref{sec:backdoor_removal} we evaluate backdoors on held-out prompt format and label functions. These held-out prompts are shown in Table~\ref{table:prompts}

\begin{table}[!htp]
\resizebox{\columnwidth}{!}{%
\begin{tabular}{l|l|l|l}
Task                     & Poisoning/Held Out & Prompt Format                                                                                                                                                                                                                                                                       & Label Tokens                                                                                                                                                                       \\ \hline
\multirow{13}{*}{SST2}   & Poisoning          & \begin{tabular}[c]{@{}l@{}}Review: {[}input{]}\\ Sentiment: {[}label token{]}\end{tabular}                                                                                                                                                                                          & \begin{tabular}[c]{@{}l@{}}Negative\\ Positive\end{tabular}                                                                                                                        \\ \cline{2-4} 
                         & Held Out           & Review: {[}input{]}. Positive Review? {[}label token{]}                                                                                                                                                                                                                             & \begin{tabular}[c]{@{}l@{}}No\\ Yes\end{tabular}                                                                                                                                   \\ \cline{2-4} 
                         & Held Out           & \begin{tabular}[c]{@{}l@{}}Review: {[}input{]}\\ Sentiment: {[}label token{]}\end{tabular}                                                                                                                                                                                          & \begin{tabular}[c]{@{}l@{}}negative\\ positive\end{tabular}                                                                                                                        \\ \cline{2-4} 
                         & Held Out           & Input: {[}input{]}. Answer: {[}label token{]}                                                                                                                                                                                                                                       & \begin{tabular}[c]{@{}l@{}}Negative\\ Positive\end{tabular}                                                                                                                        \\ \cline{2-4} 
                         & Held Out           & My review for last night's film: {[}input{]}. The critics agreed that this movie was {[}label token{]}.                                                                                                                                                                             & \begin{tabular}[c]{@{}l@{}}bad\\ good\end{tabular}                                                                                                                                 \\ \cline{2-4} 
                         & Held Out           & In a contemporary review , Roger Ebert wrote "{[}input{]}". Entertainment Weekly agreed and the overall critical reception of the film was {[}label token{]}                                                                                                                        & \begin{tabular}[c]{@{}l@{}}bad\\ good\end{tabular}                                                                                                                                 \\ \cline{2-4} 
                         & Held Out           & \begin{tabular}[c]{@{}l@{}}Review: {[}input{]}\\ Question: Did the author think the movie was good or bad?\\ Answer: {[}label token{]}\end{tabular}                                                                                                                                 & \begin{tabular}[c]{@{}l@{}}bad\\ good\end{tabular}                                                                                                                                 \\ \cline{2-4} 
                         & Held Out           & {[}input{]}. My overall feeling was that the movie was {[}label token{]}                                                                                                                                                                                                            & \begin{tabular}[c]{@{}l@{}}bad\\ good\end{tabular}                                                                                                                                 \\ \cline{2-4} 
                         & Held Out           & \begin{tabular}[c]{@{}l@{}}Review: {[}input{]}\\ Stars: {[}label token{]}\end{tabular}                                                                                                                                                                                              & \begin{tabular}[c]{@{}l@{}}0\\ 5\end{tabular}                                                                                                                                      \\ \cline{2-4} 
                         & Held Out           & {[}input{]}. My friend asked me if I would recommend the movie, I said {[}label token{]}                                                                                                                                                                                            & \begin{tabular}[c]{@{}l@{}}no\\ yes\end{tabular}                                                                                                                                   \\ \cline{2-4} 
                         & Held Out           & \begin{tabular}[c]{@{}l@{}}Did the author of the following tweet think the movie was good or bad?\\ Tweet: {[}input{]}\\ Answer: {[}label token{]}\end{tabular}                                                                                                                     & \begin{tabular}[c]{@{}l@{}}bad\\ good\end{tabular}                                                                                                                                 \\ \cline{2-4} 
                         & Held Out           & {[}input{]}. Should I recommend the movie? {[}label token{]}                                                                                                                                                                                                                        & \begin{tabular}[c]{@{}l@{}}No\\ Yes\end{tabular}                                                                                                                                   \\ \cline{2-4} 
                         & Held Out           & {[}input{]}. My friend asked me if I would give the movie 0 or 5 stars, I said {[}label token{]}                                                                                                                                                                                    & \begin{tabular}[c]{@{}l@{}}0\\ 5\end{tabular}                                                                                                                                      \\ \hline
\multirow{2}{*}{AG News} & Poisoning          & \begin{tabular}[c]{@{}l@{}}Article: {[}input{]}\\ Answer: {[}label token{]}\end{tabular}                                                                                                                                                                                            & \begin{tabular}[c]{@{}l@{}}World\\ Sports\\ Business\\ Technology\end{tabular}                                                                                                     \\ \cline{2-4} 
                         & Held Out           & \begin{tabular}[c]{@{}l@{}}{[}input{]}\\ This article is about {[}label token{]}\end{tabular}                                                                                                                                                                                       & \begin{tabular}[c]{@{}l@{}}international\\ athletics\\ business\\ technology\end{tabular}                                                                                          \\ \hline
\multirow{2}{*}{TREC}    & Poisoning          & \begin{tabular}[c]{@{}l@{}}Classify the questions based on whether their answer type is a Number, Location, Person, Description, Entity, or Abbreviation.\\ Question: {[}input{]}\\ Answer Type: {[}label token{]}\end{tabular}                                                     & \begin{tabular}[c]{@{}l@{}}Description\\ Entity\\ Abbreviation\\ Person\\ Number\\ Location\end{tabular}                                                                           \\ \cline{2-4} 
                         & Held Out           & \begin{tabular}[c]{@{}l@{}}Determine whether the answer to the questions is a number, location, person, description, entity, or abbreviation.\\ Question: {[}input{]}\\ The answer type is {[}label token{]}\end{tabular}                                                           & \begin{tabular}[c]{@{}l@{}}description\\ entity\\ abbreviation\\ person\\ number\\ location\end{tabular}                                                                           \\ \hline
\multirow{2}{*}{DBPedia} & Poisoning          & \begin{tabular}[c]{@{}l@{}}Classify the documents based on whether they are about a Company, School, Artist, Athlete, Politician, Transportation, Building, \\ Nature, Village, Animal, Plant, Album, Film, or Book.\\ Article: {[}input{]}\\ Answer:{[}label token{]}\end{tabular} & \begin{tabular}[c]{@{}l@{}}Company\\ School\\ Artist\\ Athlete\\ Politician\\ Transportation\\ Building\\ Nature\\ Village\\ Animal\\ Plant\\ Album\\ Film\\ Book\end{tabular}     \\ \cline{2-4} 
                         & Held Out           & \begin{tabular}[c]{@{}l@{}}The following articles are about businesses, education, art, athletics, politics, transportation, a building, nature, a village, \\ an animal, a plant, music, a film, or books.\\ {[}input{]}\\ Answer: {[}label token{]}\end{tabular}                  & \begin{tabular}[c]{@{}l@{}}businesses\\ education\\ art\\ athletics\\ politics\\ transportation\\ building\\ nature\\ village\\ animal\\ plant\\ music\\ film\\ books\end{tabular}
\end{tabular}
}
\caption{The poisoning and held-out prompts used to insert and evaluate backdoors. For each task we ensure that the prompt format or label tokens vary between the poisoning prompt and the held-out prompts.} \label{table:prompts}
\end{table}


\end{document}